%% file: main.tex
\documentclass[5p,number, sort&compress, final]{elsarticle}

\usepackage[utf8]{inputenc}
\usepackage{graphicx}
\usepackage[english]{babel}

\usepackage{booktabs}

\usepackage{amsmath}
\usepackage{amssymb}
\renewcommand{\vec}[1]{\boldsymbol{#1} }
\renewcommand{\matrix}[1]{\boldsymbol{\mathcal{#1}}}

\usepackage{hyperref}
\hypersetup{
	pdftitle={PARSEC},
	pdfauthor={Tobias Winchen, <tobias.winchen@physik.rwth-aachen.de>},
	pdfproducer={vim},
	bookmarks=true,
	colorlinks=false,
	hidelinks=true
}

\begin{document}
\title{PARSEC: A Parametrized Simulation Engine for Ultra-High Energy
Cosmic Ray Protons}

\author[RWTH]{Hans-Peter Bretz}
\author[RWTH]{Martin Erdmann}
\author[Hamburg]{Peter Schiffer}
\author[RWTH]{David Walz}
\author[RWTH]{Tobias Winchen\corref{cor1}}
\ead{winchen@physik.rwth-aachen.de}
\address[RWTH]{III. Physikalisches Institut A, RWTH Aachen University, Germany}
\address[Hamburg]{II. Institut für Theoretische Physik, Universität
Hamburg, Germany}

\cortext[cor1]{Corresponding author}

\begin{abstract}
We present a new simulation engine for fast generation of ultra-high
energy cosmic ray data based on parametrizations of common assumptions
of UHECR origin and propagation.  Implemented are deflections in
unstructured turbulent extragalactic fields, energy losses for protons
due to photo--pion production and electron--pair production, as well as
effects from the expansion of the universe.  Additionally, a simple
model to estimate propagation effects from iron nuclei is included.
Deflections in the Galactic magnetic field are included using a matrix
approach with precalculated lenses generated from backtracked cosmic
rays. The PARSEC program is based on object oriented programming
paradigms enabling users to extend the implemented models and is
steerable with a graphical user interface. 
\end{abstract}

\maketitle

\input{introduction}

\input{methods}

\input{results}
\input{conclusion}
\input{acknowledgements}

\appendix
\input{appendix}



\end{document}

%% file: introduction.tex
\section{Introduction}
\label{introduction}
Recent results of the Pierre Auger Observatory imply that the ultra-high
energy cosmic rays (UHECR) are accelerated at extragalactic sources with
a distribution following the large scale structure~\cite{PAO2007,
PAO2008d}. Nevertheless, the exact origin of these cosmic rays remains so
far unknown.  UHECR are likely charged particles~\cite{PAO2008d,PAO2009}
and thus deflected in the extragalactic and Galactic magnetic field.  They
can therefore be considered messengers of cosmic accelerators as well
as fields they are subjected to while propagating to the Earth.  During
their propagation from the sources to the Earth UHECR lose energy due to
interactions with different photon backgrounds. In the case of protons
the dominant processes are photo-pion and electron-pair production,
while for nuclei also photo-disintegration has to be accounted for. The
composition of UHECR is  still under
debate~\cite{HiRes2005,PAO2010}.

Progress in the understanding of origin and propagation of UHECR is
enabled by comparison of measured data with simulated UHECR data.  A
second field of application for simulated UHECR data is development and
benchmarking of observables suited for disentangling specific aspects
from the observed data.   

The interpretation of cosmic ray data relies on the consideration of 
models for the:
\begin{enumerate}
	\item  Locations and luminosities of likely sources,
	\item  Deflections in extragalactic magnetic fields,
	\item  Composition, and hence energy losses, for the UHECRs, and
	\item  Deflections in the Galactic magnetic field.
\end{enumerate}
To test these models a fast generation of sufficiently large data samples
at every point of the parameter-space is required.

The most obvious approach for a UHECR Monte-Carlo generator allowing
detailed simulations of individual phys\-ics pro\-cess\-es is to follow the
trajectories of individual particles from the source to the observer and
account for energy losses and deflections during the propagation. This
forward-propagation is for example implemented in the CRPropa 
program~\cite{Armengaud2007}.  In this approach the, compared to intergalactic
distances, small size of the observer makes the generation of large
Monte-Carlo data sets, needed for extensive parameter scans, highly
challenging. The challenge becomes more and more complex for contributions from distant sources. 
However, it can be approached by combining forward propagation codes with parametrized 
simulation software as e.g. presented here; a first attempt of such a combination 
using CRPropa and PARSEC has been presented in
reference~\cite{Dolag2011}.

A different approach aiming at UHECR mass production is to backtrack particles
starting at the observer and associate them to the objects in the source
model. This approach is well understood and documented in the
literature (e.g.~\cite{Yoshiguchi2003, Sutherland2010}). However, it
requires huge trajectory databases (e.g.~\cite{Takami2006}) and the
trajectories are associated to sources only at the end of the
simulation. 

To enable UHECR simulations for arbitrary source models with sufficient
speed for extensive parameter scans we developed PARSEC as a
parametrized simulation engine.  PARSEC calculates the probability to
observe a particle with energy $E$ from a direction with spherical
coordinates $(\phi, \theta$) by summing up the individual contributions
of point sources. It uses parameterizations for  the energy losses and
the effects from deflections in extragalactic magnetic fields to
calculate these contributions. Pre-calculated models for the Galactic
magnetic field can then be applied in a separate simulation step using
matrix techniques.  This allows for independent combination of models
for extragalactic and Galactic propagation without redundant
calculations.  From the resulting probability densities, simulated data
sets can be generated quickly. 

This publication is structured as follows: The implemented models and
details of the implementation are presented in section~\ref{methods}.
The resulting energy spectra, particle horizons,
mean deflections, and exemplary probability density maps are 
presented together with performance benchmarks in section~\ref{results}.
The conclusions in section~\ref{conclusions} are followed
by an appendix explaining uncertainties in the generation of the
galactic lenses and optional improvements concerning the computing time.

%% file: methods.tex
\section{Methods}
\label{methods}
We calculate the probability to observe a particle with energy E from
direction $(\phi, \theta)$ in discrete energy bins $E_i^l \leq E_i <
E_i^r$ and discrete directions (pixel) indexed with $j$.  For every
energy range $i$, the probabilities to observe a particle from
all directions $j$ forms a vector $\vec{p}^{obs}_i =
(p^{obs}_{i,1},p^{obs}_{i,2}, \ldots)^T$.  The total probability
distribution is thus a set of vectors $\mathcal{P}^{obs} = \lbrace
\vec{p}^{obs}_1,\vec{p}^{obs}_2, \ldots  \rbrace$ with one vector for
every energy range.  First, probability vectors $\vec{p}^{eg}_i$ for the
expected `extragalactic' distribution including all effects except the
Galactic magnetic field are calculated. In a second step this
probability vector is transformed by a matrix $\matrix{L}_i$ describing
the effects of the galactic magnetic field in energy bin $i$ reading
\begin{equation}
	 \matrix{L}_i \cdot \vec{p}^{eg}_i = \vec{p}^{obs}_i
	\label{DeflectionEquation}.
\end{equation}

\subsection{Extragalactic Propagation}
The discrete probability distribution $\vec{p}^{eg}_i$ is calculated by the sum of the
contributions of every individual source $S_k$ of a source model
$\lbrace S_1 \cdots S_N \rbrace$ to every pixel $p_{i,j}$.
We separate the contribution of $S_k$ to
$p_{i,j}$ into three factors: 
\begin{enumerate}
	\item  A factor $f^E = f^E(\gamma, z_g, B, \Lambda)$ describing the
		influence of the source spectra with spectral index $\gamma$ and
		energy loss of the particles. The energy loss depends on the red
		shift at time of injection of the particle $z_g$, and the 
elongation of the trajectory of the particle by the magnetic field with strength $B$ and
coherence length $\Lambda$.
\item A factor $f^B = f^B(B, \Lambda, D_k)$ describing the distribution
	of the flux from a source in distance $D_k$
	on multiple pixels caused by the magnetic field, and
\item A factor $f^S = f^S(D_k, L_k, B, \Lambda)$ describing the density
	of particles from source $k$ with luminosity $L_k$ after propagation
	through the extragalactic magnetic field.
	\end{enumerate}
With these ingredients the probability to observe a particle with energy
$E_i$ in pixel $j$ can be written as 
\begin{equation}
	p_{i,j} = \Gamma_i \cdot \sum_k f^E f^B
  f^S \label{sumExtragalacticPropagation}
\end{equation} 
where $\Gamma$ denotes a normalization factor ensuring $\sum_{i,j}
p_{i,j} = 1$.

\subsubsection{Energy Losses}
A particle emitted by source $S_k$ with the injected energy $E^{inj}$ propagated
a distance $c \tau$ within the extragalactic magnetic field model, and is then observed with energy $E_i$.
Here $c$ is the vacuum speed of light and $\tau$ denotes the propagation
time of the particle.
The
probability to observe the particle therefore depends on the source spectra,
the energy loss of the particle and the propagation distance, which is summarized in
$f^E$. For source spectra following a
power law described with spectral index $\gamma$ this corresponds to
\begin{equation}
	f^E = \frac{1}{1+z_g}
	\left( (E_i^{inj,r})^{\gamma+1} - (E_i^{inj,l})^{\gamma+1} \right)
	\label{sourceSpectralFactor}
\end{equation}
where the scaling factor $(1+z_g)^{-1}$ of the universe at the cosmological epoch of
particle injection accounts for cosmological time dilation.
The range of injection energies $E^{inj,l/r}_i$
is calculated with the continuous energy loss approximation by
numerically integrating
\begin{equation}
	\left(- \frac{1}{E} \frac{dE}{dx}\right) = \frac{1}{L(z,E)}
	\label{continuousEnergyLossIntegral}
\end{equation}
with the total energy loss length defined as $L(E,z)^{-1} = L_{ad}(z)^{-1} +
L_{ph}(E,z)^{-1}$. Here $L_{ad}(z) = c/H(z)$ is the energy loss
length for the adiabatic energy
loss by the expansion of the universe with Hubble parameter $H(z) = H_0
\sqrt{\Omega_{m,0} (1+z)^3 + \Omega_{\Lambda,0}}$ at the cosmological epoch $z$.
Here we use for the current matter density $\Omega_{m,0} =0.3$ and the
current dark energy density $\Omega_{\Lambda,0} =
0.7$~\cite{Spergel2003}.

For the energy loss lengths for interactions in background
photon fields $L_{ph} (E, z=0)$ as published 
in~\cite{Protheroe1996,Berezinsky2006} are implemented. Energy loss
lengths of alternative models can be easily added. As an extension, we also implemented a
simplistic model for the observed UHECR being iron nuclei. For this we
also use a continuous energy loss approximation with an attenuation
length given by the maximum of the nuclei calculated in reference~\cite{Hooper2007}. Here the iron nuclei 
do not disintegrate but keep a chargenumber of $Z = 26$. The propagation
of secondary particles is not included. As the cosmic rays have maximum
range and deflections, this model yields a maximum isotropy.
From the energy loss length of photon interactions at redshift $z=0$ 
the energy loss length at $z$ is derived from 
scaling using 
\begin{equation}
	L_{ph} (E, z) =  (1+z)^{-3} L_{ph}\left( (1+z) E, z = 0 \right)
	\label{photonEnergyL:ossLEngthScaling}
\end{equation}
to account for the increase of the energy and density of the CMB
background photons.

\subsubsection{Scattering around Sources}
The effects of the extragalactic magnetic field  are pa\-ram\-e\-trized
assuming  a turbulent field in which the
particles perform a random walk.
In a turbulent magnetic field, the flux of a single source
$S_k$ is distributed over several pixels $p_{i,j}$. 
If the particles perform a random walk,
the distribution of the angles $\alpha_{j,k}$ between the direction of source $S_k$ and
the center of pixel $p_{i,j}$
follows a Fisher distribution~\cite{Fisher1953}, which can be regarded
as normal distribution on a sphere. The second factor $f^B$
of eq.~\ref{sumExtragalacticPropagation} thus reads 

\begin{equation}
	f^B(\alpha_{j,k},\kappa) = \frac{\kappa}{4\pi \sinh{(\kappa)}} e^{(\kappa\cdot
	\cos{\alpha_{j,k})}}
	\label{FisherDistribution}
\end{equation}
with the concentration parameter $\kappa$.

The concentration parameter $\kappa$ is related to the
root mean square (rms) $\sigma$ of the deflection for small angles  by $\kappa
= 1/\sigma^2$~\cite{Mardia1972}. For small angles the root mean square of the deflection
of UHECR with charge  $Z$ in units of the electron charge $e$ in turbulent
fields with $D_k \gg \Lambda$ can be parametrized~\cite{Harari2002} as 
\begin{equation}
	\sigma = \frac{37.5^\circ}{\sqrt{3}} \sqrt{\frac{D_k}{\Lambda}} \left(
	\frac{\Lambda}{\mathrm{Mpc}} \right)
	\left(\frac{B}{\mathrm{nG}} \right)
	\left(\frac{Z}{e} \right)
	\left(\frac{E}{\mathrm{EeV}} \right)^{-1}
	\label{sigmaField}.
\end{equation} 
Here $D_k$ is the current proper distance of $S_k$ to the observer, $B$
is the
strength of the magnetic field, and 
\begin{equation}
	\Lambda = \frac{\pi}{B^2}\int_0^\infty \frac{dk}{k} B^2(k)
\end{equation}
the correlation length of the magnetic field~\cite{Achterberg1998, Harari2002}. 
For a Kolmogorov turbulent field with spectral index $n$ and minimum,
respectively maximum, correlation length $L_{\min,\max}$ it is
\begin{equation}
	\Lambda = \frac{1}{2} L_{\max}\frac{n-1}{n} \frac{{(1 - L_{\min} /
	L_{\max})}^n}{1-{(L_{\min} / L_{\max})}^{n-1}}.
	\label{eq:CorrelationLength}
\end{equation}

This parametrization for the rms of the deflection angle
does not include energy losses. To derive a first order approximation
including energy losses, we first differentiate eq.~\ref{sigmaField} with
respect to the source distance. Using $x$ as variable for the source
distance this reads
\begin{equation}
	\frac{d\sigma}{dx} = \frac{37.5^\circ}{\sqrt{3}}
	\sqrt{\frac{\Lambda}{x}}\frac{B}{E(x)} \left( 1 - \frac{x}{E} \frac{dE}{dx} \right)
	\label{differentialSigmaField}.
\end{equation}
Assuming $(dE/dx)\cdot(x/E)$ being
small for the ultra-high energies considered here the second term of 
eq.~\ref{differentialSigmaField} can be neglected. Integrating 
eq.~\ref{differentialSigmaField} to the source distance
\begin{equation}
	\sigma =  \frac{37.5^\circ}{\sqrt{3}} \sqrt{\frac{\Lambda}{x}}B\int_0^{D_k}
	\frac{1}{E(x)} dx
\end{equation}
then yields the rms of the deflection for particles from source $S_k$.

\subsubsection{Elongation of Propagation Time}
Due to deflection in magnetic fields 
the length of the trajectory of the
particles $c\tau = D_k + r$ from sources in distance $D_k$ is elongated by an extra
distance $r$. Neglecting energy losses, the additional propagation distance $r$
originating from small 
deflections in the extragalactic magnetic field can be written as
\begin{equation}
	r = 116 \mathrm{~kpc} \left(\frac{B}{\mathrm{nG}}\right)^2
	\left(\frac{Z}{e} \right)^2
	\left(\frac{D_k}{\mathrm{Mpc}}\right)^2
	\left(\frac{E}{\mathrm{EeV}}\right)^{-2}
	\left(\frac{\Lambda}{\mathrm{Mpc}}\right)
 \label{propagationDistanceWOEnergyLoss}
\end{equation}
with definitions as in eq.~\ref{sigmaField}~\cite{Achterberg1998}. 
To account for energy losses in the extended propagation length we write
eq.~\ref{propagationDistanceWOEnergyLoss} for infinitely small $dx$ 
\begin{equation}
	dr \propto  \left( \frac{2 x B^2 \Lambda Z^2}{E^2} dx - 
	\frac{2 B^2 x^2 \Lambda Z^2}{E^3}\frac{dE}{dx}dx \right)
	\label{differentialPropagationDistance}
\end{equation} 
with $ 0 < x \leq D_k$. The proportionality constant is given by
eq.~\ref{propagationDistanceWOEnergyLoss}.
Assuming again $(dE/dx)\cdot(x/E)$ to be small, the second term of 
eq.~\ref{differentialPropagationDistance} can be neglected. The result can
be written as a Riemann sum 
\begin{equation}
	r \propto \sum_i^N \frac{B^2\Lambda Z^2}{E
	\left(x_i+\frac{(x_i-x_{i-1})}{2} \right)^2} (x_i^2 -
	x_{i-1}^2)
	\label{propagationDistanceWEnergyLoss}
\end{equation}
with $x_N = D_k$.

\subsubsection{Increase of Particle Density}
The factor $f_S$ accounts for the relative individual luminosity $L_k$ of source $S_k$
and the density of particles at the position of the observer in current proper distance $D_k$. 
If the particles
propagate on a straight line, the flux from a source is distributed on a
sphere with radius of the current proper distance $D_k$ of the source
resulting in $f_S = L_k / D_k^2$.

Nevertheless, in the presence of magnetic fields the density of UHECR is
higher compared with linear propagation. In some distance to the source all information
about the origin of the UHECR is lost, and instead of a directed random
walk the density is described according to a undirected random walk. To model
this transition we simulated the trajectories of individual UHECR from
one source in a turbulent magnetic field using the
CRPropa software~\cite{Batista2013}. Using $g = \frac{B}{E}\sqrt{\Lambda}$ 
the density of UHECR in distance $D_k$ can be approximatly described by 
\begin{equation}
	f_S = \frac{L_k}{D_k^2} \left( (1 + p_1\cdot g^2\cdot D_k) (1-T) +
	T\cdot p_2\cdot g \cdot e^{-\frac{1}{2}
	\left(\frac{D_k}{x_1}\cdot g -1\right)^2}  \right)
	\label{eq:ParticleDensity}
\end{equation}
with 
\begin{equation}
	T = \frac{1}{1+\left( \frac{D_k\cdot g}{x_t} \right)^{-s}}.
	\label{eq:TransitionFunction}
\end{equation}
and parameters as given in table~\ref{tab:FitResults}.
\begin{table}[Htb]
\centering
\caption{Results of the fit of eq.~\ref{eq:ParticleDensity} to UHECR
densities in forward simulations.}
\begin{tabular}{l r@{}l c}
	\toprule
	Parameter & Val&ue & Unit \\
	\midrule
	$p_1$  & 0 &.1 & $\mathrm{nG}^{-2}\,\mathrm{EeV}^2\,\mathrm{Mpc}^{-3/2}$ \\
	$s$    &  5& & --- \\
	$x_t$  & 70& &
	$\mathrm{nG~EeV}^{-1}\,\mathrm{Mpc}^{3/2}$\\
	$p_2$  &  9& &
	$\mathrm{nG}^{-1}\,\mathrm{EeV}\,\mathrm{Mpc}^{-1/2}$\\
	$x_1$  & 140& & $\mathrm{nG}\,\mathrm{EeV}^{-1}\,\mathrm{Mpc}^{3/2}$:\\
	\bottomrule
\end{tabular}
\label{tab:FitResults}
\end{table}
For $0.1\,\mathrm{nG~EeV}^{-1}\,\mathrm{Mpc}^{1/2} < g <
2.5\,\mathrm{nG~EeV}^{-1}\, \mathrm{Mpc}^{1/2}$ the
deviation of eq.~\ref{eq:ParticleDensity} from the simulated values is typically below 20\%.
For a homogeneous distribution of sources, the resulting observed
spectrum agrees
also within 20\% with the universal spectrum~\cite{Aloisio2004}.

\subsection{Galactic Magnetic Field}
To model particle propagation in the Galactic magnetic field we neglect
energy losses during the relatively short Galactic propagation. As there
is no random process in this model, the trajectories of all cosmic rays
with energy $E_i$ traversing the Galaxy are completely defined by the
point of entry into the Galaxy $\vec{C}$ and their direction $(\phi,
\theta)$.  The effects of the Galactic magnetic field can thus be
addressed as magnetic lensing, i.e.\ a mapping of every point of the phase space
$(\vec{C}, \theta, \phi)_{eg}$ outside the Galaxy, i.e.\ where the
effect of the Galactic magnetic field can be neglected, to a point
$(\vec{D}, \theta, \phi)_{obs}$ within the Galaxy~\cite{Harari1999, Harari2000}.

We assume that the observed spatial density of cosmic rays does not
depend strongly on the position of the observer so that Earth can be
safely modelled as point and only the directions from which a particle
are observed at the position of the Earth are relevant.  We index the discrete
observed directions (pixel) with $m$. A particle with energy $E_i$ entering the galaxy at a point
$\vec{C}$ on the surface of the galaxy at a discrete direction denoted
with index $n$ is thus always observed at Earth at direction $m$.

The sources considered here are generally in a large distance compared
to the size of the galaxy, which reduces the galaxy to a point in view
of the source. We assume that the spatial density of cosmic rays from
one source on the edge of the Galaxy, i.e. the sphere which encloses the
volume for which the Galactic magnetic field is considered, is governed
by geometry and not by effects from deflection in the extragalactic
magnetic field.  The directions of entry $n$ for particles with energies
$E_i$ can therefore be averaged over all points of entry $\vec{C}$.
Thus, and due to the limited number of pixels,  a particle with energy
$E_i$ entering the galaxy from direction $n$ can be deflected into
several observed directions $m$.

The probability of observing a particle on Earth from direction $m$ which entered
the galaxy from direction $n$ is $l_{m,n}$. The $l_{m,n}$ form a matrix
$\matrix{L}_i$ which represents the galactic lens for energy $E_i$. 
The model for the Galactic magnetic field is completely described by a set of
matrices $\lbrace \matrix{L}_1 \cdots \matrix{L}_N
\rbrace$ with the energy index $i = 1\ldots N$. 

\begin{figure}[tb]
	\centering
		\includegraphics[width=\columnwidth]{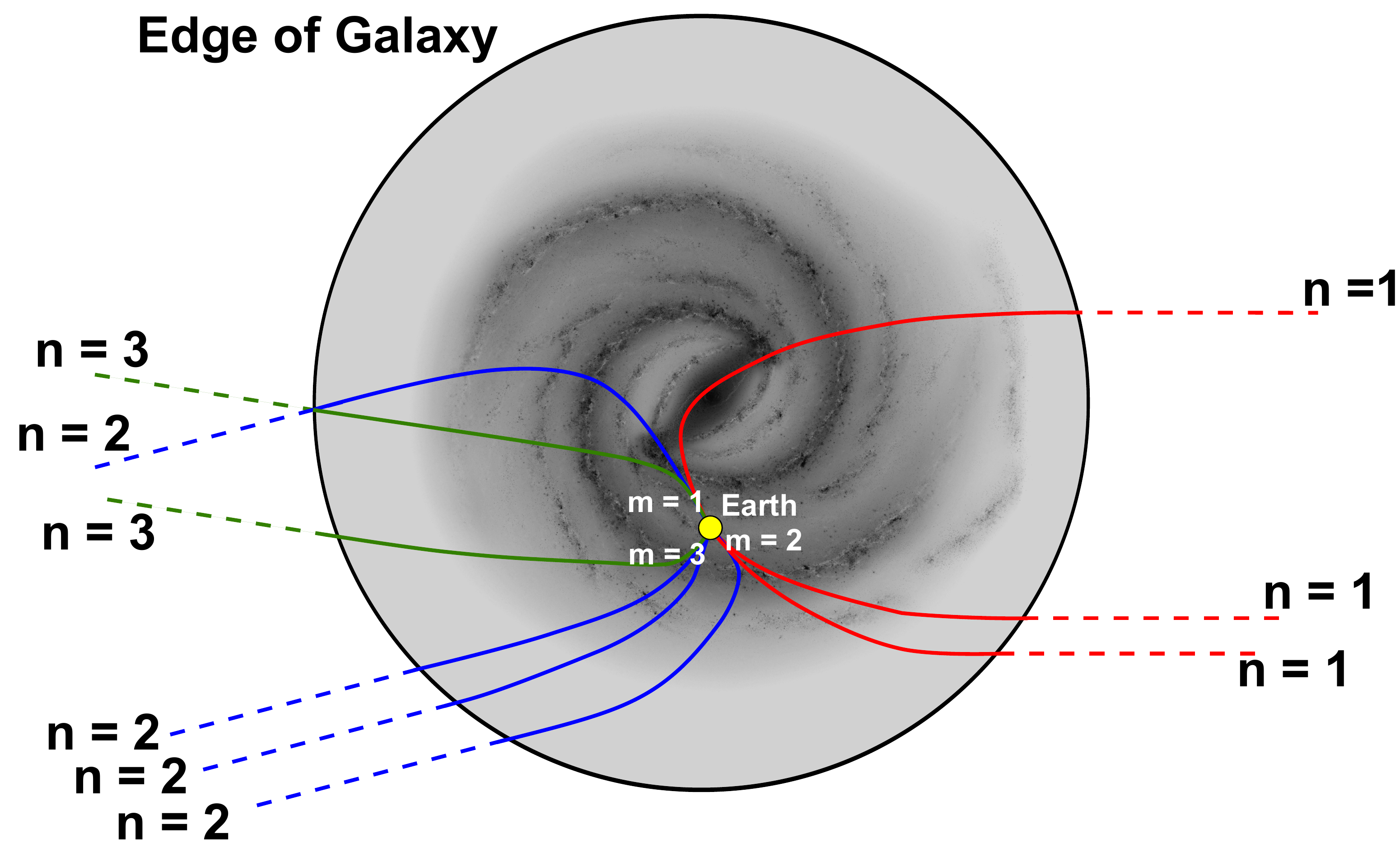}
	\caption{Sketch of the creation of a galactic lens from backtracking
	data. Cosmic rays are emitted from earth in three directions indexed
with $m$. The directions in which the rays leave the galaxy are indexed
with $n$.}\label{fig:GalacticLensSketch}
\end{figure}

The individual matrices $\matrix{L}_i$ can be generated by backtracking cosmic
rays with isotropic starting directions from the earth with the following technique. The starting directions of
backtracked particles are
binned in $N$ pixels indexed by $m$. The directions in which the cosmic rays
leave the galaxy are binned into $N$ pixels indexed by $n$. Counting all
trajectories leads to a matrix $\matrix{\tilde{L}}_i$ with elements
$\tilde{l}_{m,n}$.
For three directions and nine backtracked particles the procedure is
illustrated in figure~\ref{fig:GalacticLensSketch}. 
We normalize 
$\matrix{\tilde{L}}_i$ by the maximum of unity norms $\Vert \matrix{\tilde{L}}_i \Vert_1$ of all lenses
reading 
\begin{equation}
	\matrix{L}_i = \frac{1}{\max{\Vert \matrix{\tilde{L}}_i \Vert_1}} \matrix{\tilde{L}}_i
	\label{MatrixMaxNorm}.
\end{equation} 
Each element $l_{m,n}$ of $\matrix{L}_i$ is the probability that a particle entering
the galaxy in pixel $n$ is observed in direction $m$.

As a consequence of Liouville's theorem, that the phase space along
trajectories that satisfy the Hamiltonian equations is constant, an
isotropic distribution of cosmic rays outside the galaxy is observed as
an isotropic distribution at any point of the galaxy~\cite{Lemaitre1933,
Swann1933, Harari2000}. This important property of the Galactic magnetic field is
correctly modeled by this technique, if the directions $m$ are uniformly
sampled in the backtracking, respectively the unity norm $\Vert.\Vert_1$
of all row vectors $\vec{l}_m$ are identical.

In general the Galactic magnetic field modifies the energy spectrum of cosmic
rays depending on the positions of the sources $S_k$ as the flux from individual regions in the
sky is suppressed or enhanced~\cite{Harari2000}.
While the relative
deformation of the energy spectrum is accounted for in the normalization procedure
described above, no information about a suppression of the
total flux by the Galactic magnetic field is obtained by the generation
of the galactic lenses from backtracking. 

Galactic lenses, generated from backtracking Monte-Carlo data in the
described way, introduce an uncertainty in the observed probability
distribution. This uncertainty is discussed in~\ref{LensUncertainty}.

\subsection{Technical Realization}
PARSEC is implemented  as C++ code with a Python interface. It is based
on the Physics Extension Library (PXL)~\cite{Brodski2009}.
PXL is a
collection of C++ libraries with a Python interface providing classes and
templates for experiment independent high-level physics analysis. The
usage of the PXL libraries facilitates modular object-oriented
programming and allows graphical steering of the simulation components
using the VISPA program~\cite{Bretz2012a}. 

The individual simulation steps are implemented as separate PXL modules
which can
be individually connected and configured to a simulation chain using the
graphical user interface (GUI) of VISPA. A  realization of a UHECR
scenario is
represented by a data container, which is consecutively processed by the 
following modules.

\subsubsection*{Source Model}
Sources of UHECR are represented as individual objects. They are added
to the realization with user-defined coordinates in a Python or C++
module. An exemplary module for isotropic source distributions is
included in PARSEC. Modules generating sources e.g.\ from astronomical
catalogues can be created by users.

\subsubsection*{Extragalactic Field Model}
From the sources in the simulation the
probability vectors for extragalactic propagation are calculated for
a user-defined discretization of the energies and directions. The
calculation is separated into C++ classes for the propagation
and the energy loss, each based on an abstract interface. The abstract
interfaces are implemented as subtypes for the random-walk propagation in
turbulent fields, respectively the described energy loss for proton and
iron UHECR. This polymorphic design enables users to modify and
extend the individual components independently.

\subsubsection*{Galactic Field Model}
 For an angular resolution of the discretization better than
 $\approx 1^{\circ}$ matrices of about $50 000 \times 50 000$ elements
 are needed. However, as in typical Galactic magnetic field models particles from
 most directions are not distributed over the whole sky, the
 matrices $\matrix{L}_i$ are only sparsely populated.
The lenses for the Galactic field are consequently implemented using a common linear
algebra library which features sparse matrices~\cite{UBLAS}. This
enables calculation of eq.~\ref{DeflectionEquation} with reasonable
consumption of resources. 
PARSEC includes tools for generation of the lenses 
from backtracking data from the CRT~\cite{Sutherland2010} and
CRPropa~\cite{Batista2013} programs.

The galactic lenses are independent of the PARSEC module for the
extragalactic propagation and can be used to calculate the deflection of individual
cosmic rays. Spline interpolation and numeric integration routines used
in the program are taken from the GNU Scientific Library~\cite{Galassi2009}.

%% file: results.tex
\section{Results}
\label{results}

\subsection{Energy spectrum}
\begin{figure}[tb]
	\centering
		\includegraphics[width=.99\columnwidth]{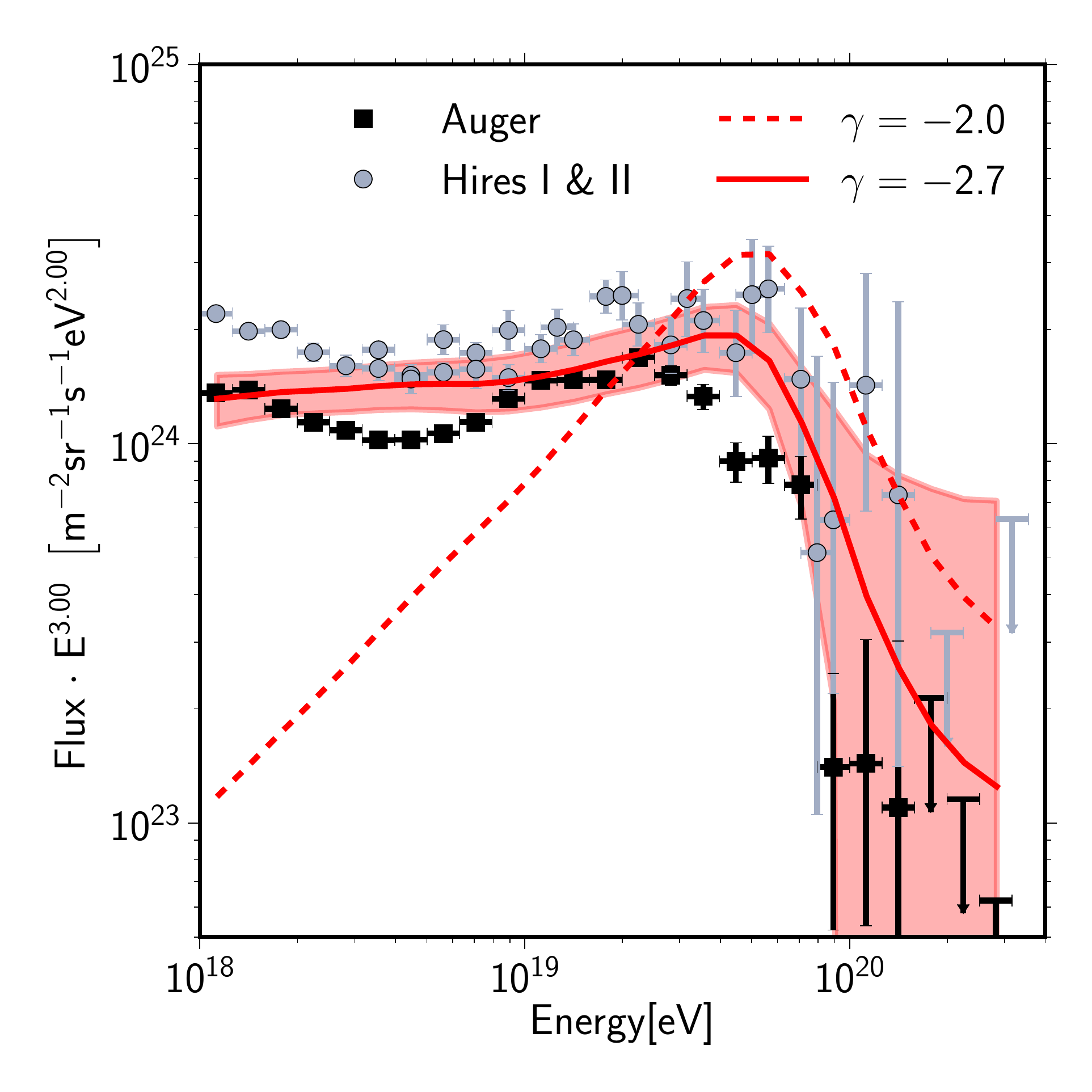}
	\caption{Energy spectra of UHECR generated by PARSEC and observed
	spectra reported by the Pierre Auger Observatory~\cite{PAO2010a} and
	the HiRes experiment~\cite{HiRes2008}.  The simulation results are obtained
	from sources with spectral indices $\gamma = -2.7$
	and $\gamma = -2.0$ in an extragalactic magnetic field with $B=1$ nG
	and $\Lambda = 1$ Mpc.  The shaded regions indicate the spread in the
	simulated realizations.}
  \label{fig:Spectra}
\end{figure}
The energy spectrum of cosmic rays is a key distribution for comparisons
of models with observations. From PARSEC simulations the energy spectra
$dN/dE(E_i)$ are calculated using $dN/dE(E_i) = \hat{L} \cdot \Vert
\vec{p}_i \Vert_1$ with normalization factor $\hat{L} = L_0 / \sum_i^N
L_i$ where the luminosity scale $L_0$ fitted to data or set by a source
model.

In Figure~\ref{fig:Spectra} energy spectra obtained with the PARSEC
program are compared with the observed energy spectra of the Pierre
Auger Observatory~\cite{PAO2010a} and HiRes experiment~\cite{HiRes2008}.
The spread of the energy spectra of 50 realizations with different
source positions and the mean energy spectrum are shown for two different
spectral indices $\gamma$ of the injection spectra.  The simulated
spectra were generated using isotropically distributed sources with a
density of $10^{-5}$ Mpc$^{-3}$ and an extragalactic field of strength
$B = 1$ nG with correlation length $\Lambda = 1$ Mpc. The probability
maps $\vec{p}_i$ have been calculated for 100 log-linear spaced energies
$E_i$ from $10^{18.5}$ eV to $10^{20.5}$ eV in the simulation.  The
normalization factor $\hat{L}$ has been fitted to match the result from the
Pierre Auger Collaboration at an energy of 22 EeV.

\subsection{Particle horizons}
\begin{figure}[tb]
	\centering
		\includegraphics[width=.99\columnwidth]{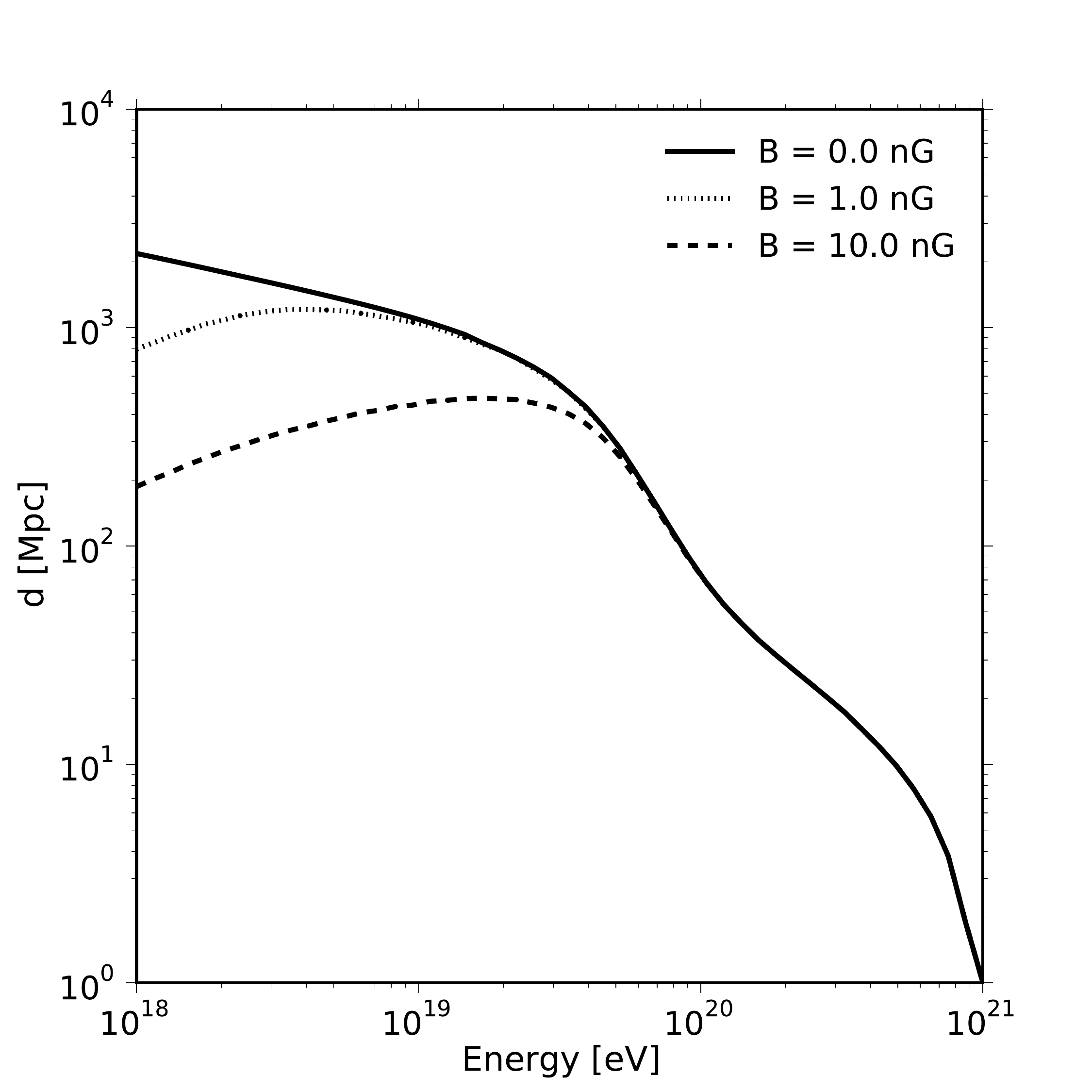}
	\caption{Horizon $d$ for protons with observed energy $E$. The
	curves show $d$ for several strengths of a turbulent magnetic field
	$B$ with
	correlation length $\Lambda = 1$ Mpc.}
	\label{ParticleHorizon}
\end{figure}
From the maximum injection energy $E_{max}$ and the model of the energy
loss a particle horizon can be derived which corresponds to the maximum
linear distance a particle can originate from. Following 
eq.~\ref{propagationDistanceWEnergyLoss} this horizon depends on the
extragalactic magnetic field model and the observed particle energy.  In
Figure~\ref{ParticleHorizon} the horizon for different field strengths
is shown as a function of the energy for sources with a maximum
injection energy of $E_{max} = 1000$ EeV.

In case of a non-zero extragalactic magnetic field, the horizon is
reduced as the trajectories are elongated. The effect is visible
below $20$ EeV for a $1$ nG field and at higher energies for stronger
fields. Without scaling the energy losses with red-shift $z$, the
horizon is not reduced for lower energetic particles, as the energy
loss only depends on the particles energy and not the time. The energy
thus monotonically increases with the linear distance.

For an energy of 100 EeV the distance from within 90\% of the UHECR flux
originates from is calculated to $d_{90}$ = 45 Mpc in case of zero
magnetic field and a spectral index of the sources $\gamma = 2.0$. This
is $\approx 30\%$ lower than the result obtained  from simulations with
CRPropa~\cite{Armengaud2007} and a known feature of the continuous energy loss approximation
\cite{Kachelriess2008}.

\subsection{Mean Deflection in Magnetic Fields}
\begin{figure*}[htb]
	\begin{center}
		\includegraphics[width=.99\textwidth]{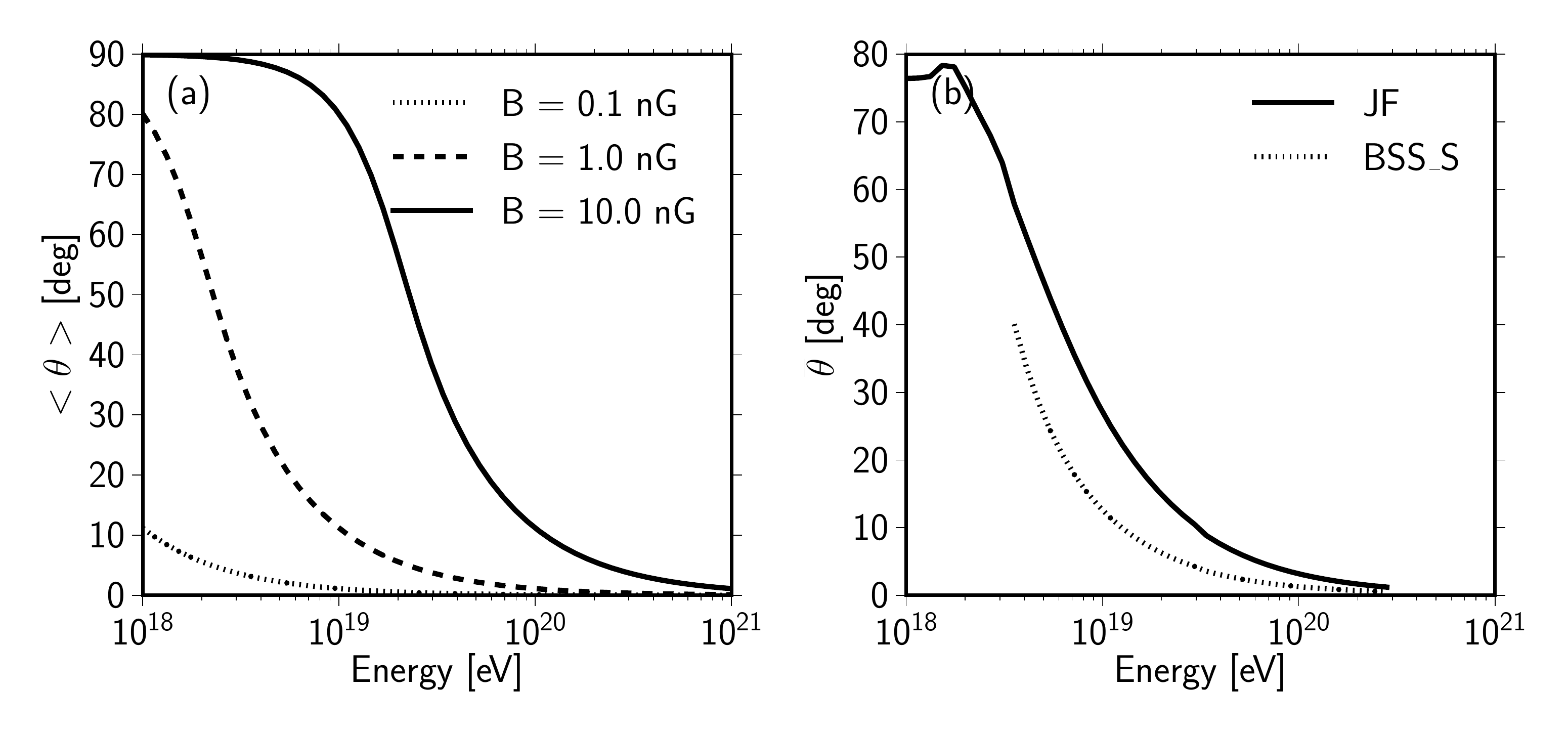}
	\end{center}
	\caption{Mean deflection of cosmic ray protons in \textbf{(a)} turbulent extragalactic
	fields and \textbf{(b)} BSS\_S  and JF Galactic magnetic field model.} 
	\label{MeanDeflection}
\end{figure*}

The expectation value for an angle $\theta$ following a  Fisher
distribution with concentration parameter $\kappa$ is calculated as 
\begin{equation}
	\langle\theta\rangle = \frac{\pi}{2 \sinh{\kappa}} \cdot ( I_0(\kappa) - e^{-\kappa})
	\label{fisherMean}
\end{equation}
with $I_0$ representing the modified Bessel function of order $0$. 
From this the mean deflection in the extragalactic field can be
calculated for the parameters of the magnetic field and source model 
using eq.~\ref{sigmaField}. 

In Figure~\ref{MeanDeflection}~a) the mean deflection of protons 
from a source at 10 Mpc distance in extragalactic magnetic fields with
three different strengths and a coherence length $\Lambda =1$ Mpc are
shown as a function of the energy. 
A mean deflection of $90^\circ$ corresponds to an isotropic
arrival distribution of the UHECR.

The mean deflections $\bar{\Theta}$ in the Galactic magnetic field can be 
directly calculated from the lens $\matrix{L}_i$ as 
\begin{equation}
	\bar{\Theta} = \frac{1}{\sum_{m,n} l_{m,n}} \sum_{m,n}
 l_{m,n}\arccos{\left( \vec{e}_m \cdot \vec{e}_n \right)}
\end{equation}
with $\vec{e}_{n}, \vec{e_{m}}$ being the unit vector in the direction of pixel $m$,
respectively $n$.  In general  the mean deflection in the Galactic
magnetic field
depends on the source configuration due to the suppression of individual
regions by the galactic lens. 
In Figure~\ref{MeanDeflection}~b) the mean deflection from lenses for
the JF model~\cite{Jansson2012} and the BSS\_S model~\cite{Stanev1997,
Harari1999} of the Galactic magnetic field
are displayed. For the BSS\_S model a field normalization of $B_0 = 0.48~\mu$G  
and scale heights of $z_1 = 0.95$ kpc and $z_2 = 4.0$ kpc were chosen.

\subsection{Exemplary Probability Distributions}
\begin{figure*}[htb]
	\centering
	\includegraphics[width=\textwidth]{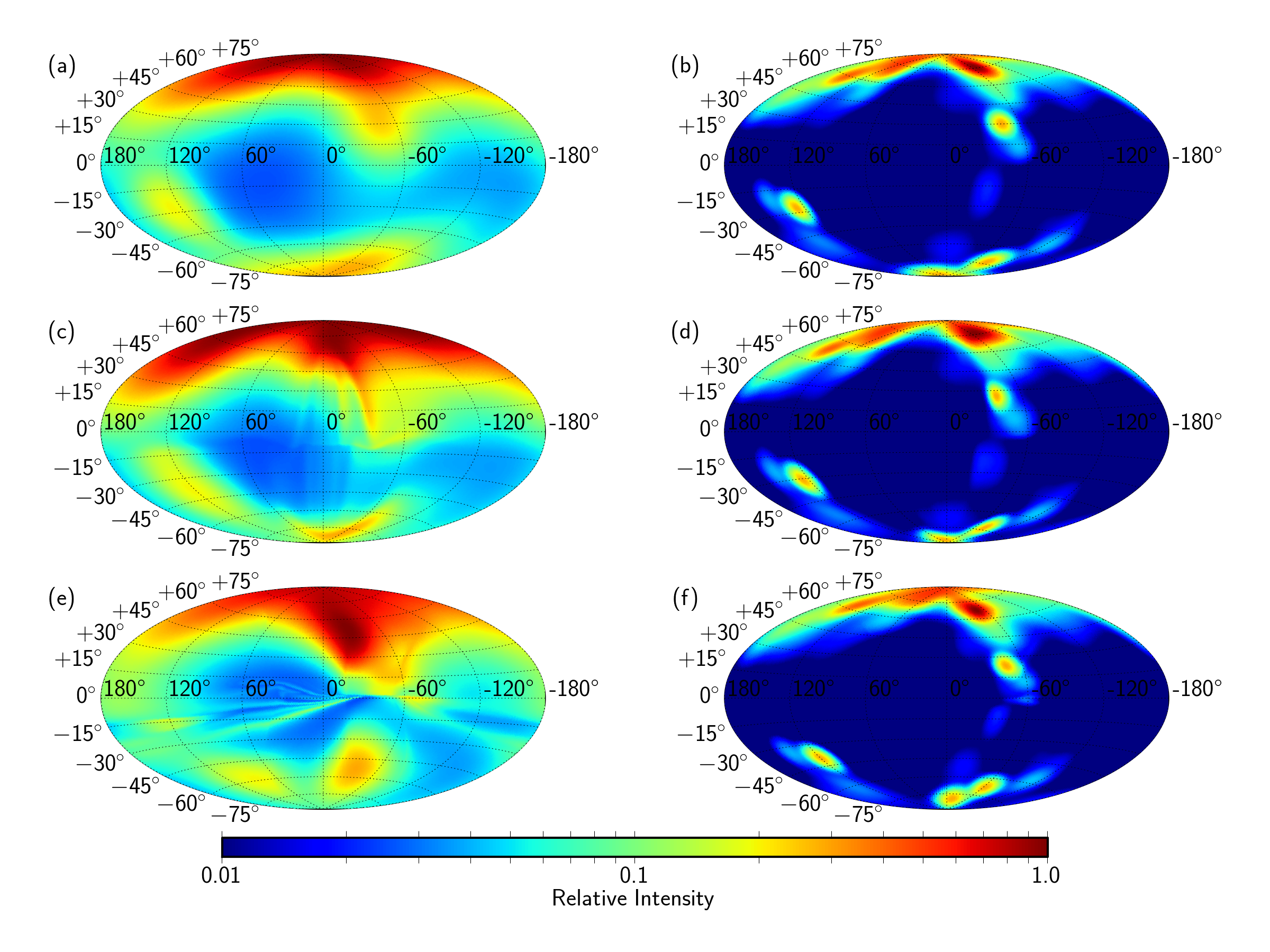}
	\caption{Hammer projected probability density maps for cosmic ray
		protons with an observed
	energy of $E=10$ EeV (left column) and $E=30$ EeV (right column) in
	galactic coordinates.
	Red indicates higher, blue lower probability on an arbitrary scale.
	From top to bottom the
	panels show: \textbf{(a,b)} Probability density before applying the galactic lens,
	\textbf{(c,d)} probability density after application of the galactic
	lens of a BSS\_S model, and \textbf{(e,f)}
	probability density after 
	application of the galactic lens for the JF2012 model.} 
	\label{ExemplarySkyMaps}
\end{figure*}

To demonstrate the capabilities of PARSEC we generated an exemplary
result with source positions taken
from the 12th edition of the catalogue of quasars and active galactic
nuclei (AGN) by Véron-Cetty and Véron~\cite{Veron-Cetty2006}. Every AGN
of the catalogue up to a distance of 1000 Mpc has been considered.
For the extragalactic field we chose a field strength $B=3$ nG and a
correlation length of $\Lambda = 1$ Mpc. The resulting probability maps
from the extragalactic propagation
$\vec{p}_{eg}^i$ are shown in the top row of Figure~\ref{ExemplarySkyMaps}
for the two different energies $E_1 = 10$~EeV (Fig.~\ref{ExemplarySkyMaps} a)
and $E_2 = 30$~EeV (Fig.~\ref{ExemplarySkyMaps}~b).

The second row of Figure~\ref{ExemplarySkyMaps} (c,d) displays the same
probability maps after application of the galactic lenses for a
BSS\_S model of the Galactic magnetic field with a normalization
of $B_0 = 0.48~\mu$G and scale heights of $z_1 = 0.95$
kpc and $z_2 = 4.0$ kpc. The lenses
have been created by backtracking $10^6$ protons with the
CRT program~\cite{Sutherland2010} for 100 log-linear spaced mono-energetic
simulations from $10^{18.5}$ eV to
$10^{20.5}$ eV. The lenses and probability vectors have been discretized
into 49,152 equal area pixels following the scheme implemented in the
HEALpix software~\cite{Gorski2005}. 
The third row of Figure~\ref{ExemplarySkyMaps} (e,f) shows the
probability maps of (a,b) after application of a lens for the JF
model~\cite{Jansson2012} of the Galactic magnetic field.

\subsection{Performance}
\begin{figure}[htb]
	\centering
		\includegraphics[width=.9\columnwidth]{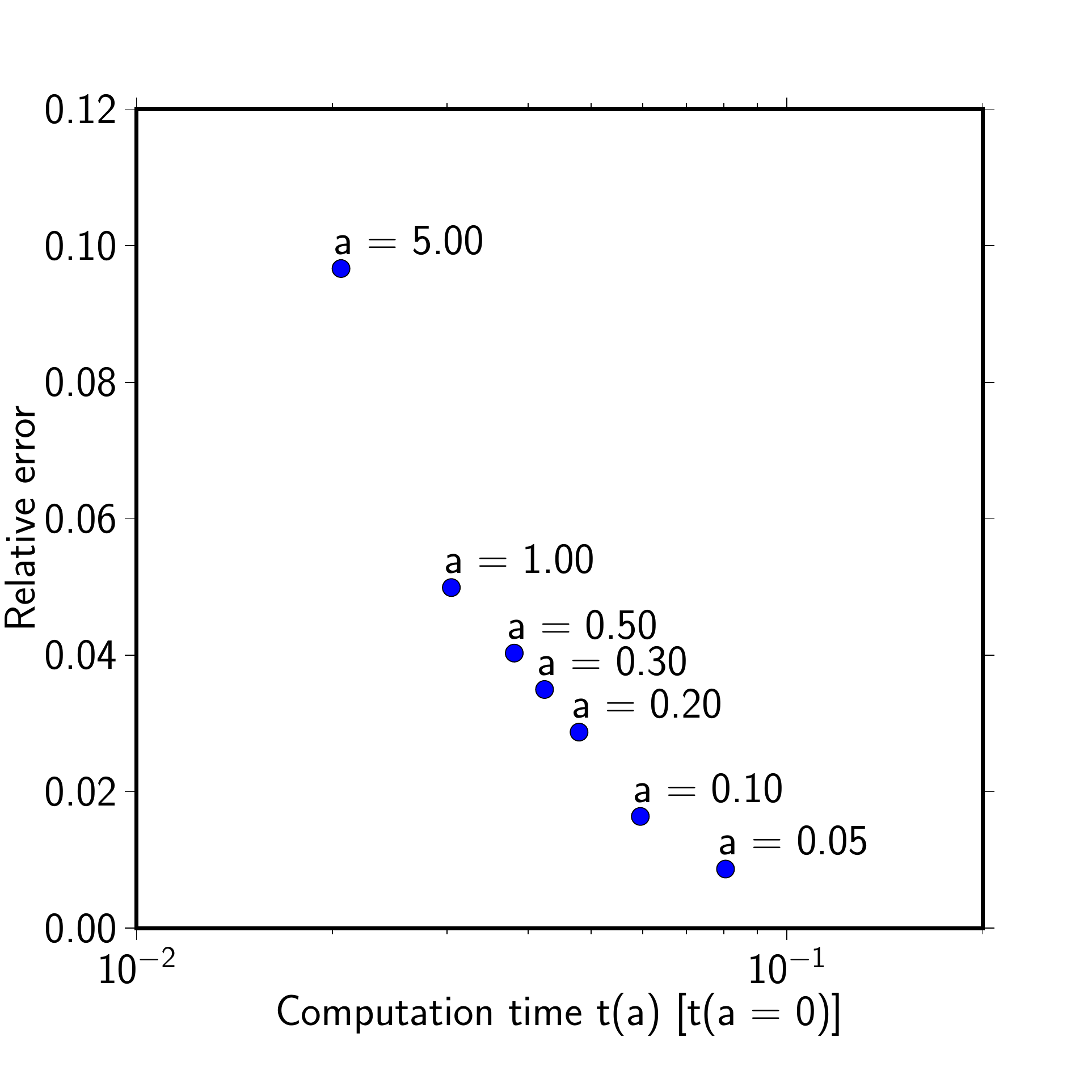}
	\caption{Computation time for the exemplary scenario and resulting error $\epsilon$
		for different choices of the cut-off parameter $a$. Times are given
	in units of the time needed for the full calculation.}
	\label{accelerationFactor}
\end{figure}

The exemplary simulation was performed in 6,690 sec using a single core of
a Lenovo Thinkpad T400 notebook with 4 GB RAM and an Intel Core 2 Duo
P8600 2.4 GHz CPU. The notebook has been benchmarked with a
SPECfp\_base2006 rate of 12.2 and SPECint\_base2006 rate of 15.1
\cite{SPEC}. Peak memory usage of the program was 0.5 GB. The size of
the BSS\_S magnetic lens on disk is 262 MB.

The performance can be improved by approximating the individual
contribution from distant sources as an i\-so\-tro\-pic component.  This approximation
introduces an error $\epsilon$ to the extragalactic probability vector $\vec{p}_{eg}^i$.  
An example realization
with isotropic source distribution with $10^{-4}$ sources per Mpc$^{3}$ up
to 1000 Mpc is used to demonstrate the effect.
In Figure~\ref{accelerationFactor} 
the resulting error $\epsilon = \Vert \vec{p}_{unbiased} -
\vec{p}_{biased} \Vert_1$ is displayed as a function of the computation
time for various values of the cut-off
parameter $a$ (see~\ref{performanceBoost}). Here $a$ represents a
measure of the strength of the anisotropic signal contribution not considered in the
calculation.
By this method the computation time can be reduced by an order of
of magnitude for $a = 0.05$, which introduces an uncertainty less than 1\%.

%% file: conclusion.tex
\section{Conclusions}
\label{conclusions}
We have presented a parametrized Monte-Carlo generator for fast creation
of simulated UHECR datasets. The code includes effects of proton
propagation through extragalactic and Galactic magnetic field models as
well as energy losses during propagation. Also implemented is an
extension to a simple model of iron propagation.  Arbitrary source
models can be easily tested with the code. The generation of datasets is
sufficiently fast to enable extensive parameter scans of the models.
The modular design of PARSEC combined with the graphical steering is
suited for user extensions of the code and implementation of user
defined models.  The source code of PARSEC as well as exemplary magnetic
lenses are available under a GNU General Public License from
\url{http://www.physik.rwth-aachen.de/parsec}.

%% file: acknowledgements.tex
\section*{Acknowledgements}
The authors gratefully thank Brian Baughman and Michael Sutherland for
mass production of cosmic rays with the CRT program which enabled the
generation of the galactic lenses, and excellent comments on the
manuscript. We also thank Patrick Younk, Haris Lyberis, and  Manlio De Domenico for
fruitful discussions, and Manuel Giffels for support in performing the
SPEC benchmarks. \\ This work is supported by the Mi\-nis\-ter\-ium
f\"ur Wis\-sen\-schaft und For\-schung, Nordrhein-Westfalen, and the
Bundes\-mi\-nis\-ter\-ium f\"ur Bil\-dung und For\-schung (BMBF). T.
Winchen gratefully acknowledges funding by the
Friedrich-Ebert-Stif\-tung.

%% file: appendix.tex
\section{Appendix}
\subsection{Uncertainties of Galactic Lenses}
\label{LensUncertainty}
From two realizations of the same model for the galactic magnetic field an upper limit
of the introduced error can be derived as follows.
For eq.~\ref{DeflectionEquation} it is $\Vert \vec{p}_{obs.} \Vert_1
\leq 1$ as  individual regions of the sky are suppressed. However, from $\matrix{L}\in \mathbb{R}^{N \times N} $
with $l_{m,n}$ we can generate 
\begin{equation}
	\matrix{\hat{L}} = \left( \begin{array}{ccc}
 l_{1,1} & \cdots & l_{1,N} \\
 \vdots& \ddots & \vdots \\
 l_{N,1}& \cdots& l_{N,N} \\
 s_{1}& \cdots& s_{N} \\
\end{array} \right)
\label{Lhat}
\end{equation}
with $\matrix{\hat{L}} \in \mathbb{R}^{N+1 \times
N}$ and $s_n = 1 - \Vert \vec{l}_{n} \Vert_1$ such that 
$\matrix{\hat{L}}\cdot \vec{p}_{eg} = \vec{\hat{p}}_o $ with
	$\vec{\hat{p}}_o \in \mathbb{R}^{N+1}$ and $\vec{\hat{p}}_o^T = ( p_1,
	\cdots p_{N}, s )$. $s_n$ represents the suppression the UHECR flux
	from the extragalactic direction $n$
	and $s = \sum s_n$ the total suppression of $\vec{p}_{eg}$  by $\matrix{L}$.
	By this definitions it is $\Vert \vec{\hat{p}}_o \Vert_1 = 1$.

	Let $\matrix{\hat{L}}_1$ be a realization
	of the `true' lens $\matrix{\hat{L}}$, then application of
	$\matrix{\hat{L}}_1$ in eq.~\ref{DeflectionEquation} introduces an
	uncertainty $\vec{\hat{\delta p}}$
	\begin{equation}
		\matrix{\hat{L}}_1\cdot \vec{p}_{eg} = \vec{\hat{p}}_o +
		\vec{\hat{\delta p}}
		\label{DeflectionEquationWithUncertainty}
	\end{equation} which depends on the extragalactic probability density
	$\vec{p}_{eg}$, or the individual configurations of the source and
	extragalactic propagation models, respectively. 
	
	If $\vec{p}_{eg}$ is known, the
	uncertainty can be calculated, as	we can approximate the true lens
	$\matrix{\hat{L}}$ by the mean of individual realizations. In the
	following calculations only two realizations
	$\matrix{\hat{L}}_1$ and $\matrix{\hat{L}}_2$ of
	$\matrix{\hat{L}}$ are used in order of clarity.
	
	For two realizations the true lens can be approximated as $\matrix{\hat{L}} =
	\frac{1}{2} (\matrix{\hat{L}}_1 + \matrix{\hat{L}}_2)$. Using this we
	substitute
	$\matrix{\hat{L}}_1$ in 
	eq.~\ref{DeflectionEquationWithUncertainty} which yields
	\begin{equation}
		\frac{1}{2} \matrix{\delta \hat{L}} \cdot \vec{p}_{eg}=
		\vec{\hat{\delta p}}
		\label{errorEstimate}
	\end{equation}
	with $\matrix{\delta \hat{L}} = \matrix{\hat{L}}_1 - \matrix{\hat{L}}_2$.

	For 
	\begin{equation}
		\vec{\hat{\delta p}} = \epsilon \cdot \vec{\hat{p}}_o
	\end{equation}
	resembling a uniform uncertainty on the sky
	and unknown $\vec{p}_{eg}$ we can estimate $\epsilon$ by applying the
	unity norm $\Vert . \Vert_1$ to eq.~\ref{errorEstimate}. Using the
	Cauchy–-Schwarz inequality this reads
	\begin{equation}
	\vert \epsilon \vert \leq \frac{1}{2} \Vert \matrix{\delta \hat{L}} \Vert_1
		\label{epsEst}
	\end{equation}
	as $\Vert \vec{p}_{eg} \Vert_1 =\Vert \vec{\hat{p}}_o \Vert_1 = 1$.

	The definition of the unity norm reads $\Vert \matrix{\hat{\delta L}}
	\Vert_1 =
	\max_n \Vert \vec{\hat{\delta l}}_n \Vert_1$ with $\vec{\hat{\delta
	l}}_n$ being the $n$-th column vector of $\matrix{\hat{\delta L}}$. 
	
	Using $\matrix{\hat{\delta L}}$ as in eq.~\ref{errorEstimate} and
	$\matrix{\hat{L}}$ as in eq.~\ref{Lhat}
	$\vec{\hat{\delta l}}_n = \left( \delta l_{1,n}, \cdots l_{N,n}, \delta s_n \right)^{T}$ 
	with $\delta l_{m,n} = l^1_{m,n} - l^2_{m,n}$ being
	the difference of the elements of the matrices $\matrix{L}_{1,2}$ and
	$\delta s_n = s^1_n - s^2_n$ being the difference of the corresponding suppression factors. 
	Thus we can write
	\begin{align}
		\Vert \vec{\hat{\delta l}_n} \Vert_1 &= \sum_m \vert l^1_{m,n} -
		l^2_{m,n} \vert + \vert  s^1_n - s^2_n \vert \\
		&= \sum_m \left\vert l^1_{m,n} - l^2_{m,n} \right\vert + \big\vert 1
		- \vert l_{m,n}^1 \vert -	1 + \vert l_{m,n}^2 \vert \big\vert \\
		&= \Vert \matrix{\delta L}_n \Vert_1 + \big\vert
		\Vert\matrix{L}_1\Vert_1 -
		\Vert\matrix{L}_2\Vert_1 \big\vert
	\end{align}
	using the definitions of the suppression factors.
	This yields 
	\begin{equation}
		\epsilon \leq \frac{1}{2} \Vert \matrix{\hat{\delta L}} \Vert_1 =
		\frac{1}{2}\max_n \left( \Vert \delta \vec{L}_n
	\Vert_1 + \big\vert \Vert \matrix{L}_1 \Vert_1 - \Vert
	\matrix{L}_2\Vert_1 \big\vert \right)
	\label{epsilonApproxFinal}
	\end{equation}
	as an upper limit of the uncertainty of the lens.

	The formalism can be extended to give an upper limit of the
	uncertainty in individual directions by substitution of the scalar
	$\epsilon$ in eq.~\ref{errorEstimate} with a diagonal matrix
	$\matrix{E}$ with
	elements $\epsilon_{m,m}$	being the relative uncertainty of the probability 
	in pixel $n$. 
	Following the same calculation steps as above this yields
	\begin{equation}
		\vec{e}_n^T \cdot \matrix{E}  \cdot\vec{\hat{p}}_o = \frac{1}{2}
		\vec{e}_m^T
	 \cdot	\matrix{\hat{\delta L}}  \cdot \vec{p}_{eg}
 \end{equation}
 with beeing the $\vec{e}_m$ unit vector in direction $m$. Consequently this transforms to 
 \begin{align}
	 \vert \epsilon_{m,m} \vert &\leq \Vert \vec{e}_m^T \cdot \matrix{E} \Vert_1 \\
	 &\leq  \frac{1}{2} \Vert \vec{e}_m^T \cdot \matrix{\hat{\delta L}}
	 \Vert_1 = \frac{1}{2} \max_n \vec{\hat{\delta l}_{m}^T} 
 \end{align}
 with row vector $\vec{\hat{\delta l}}_{m}^T$ as upper limit of the
 uncertainty in a specific direction.

 From two realizations of the lenses used in section~\ref{results} we
 found a maximum uncertainty of 23\% for a UHECR energy of $10^{18.5}$~eV
	with an typical uncertainty for individual pixels of about 2\%.
 The maximum uncertainty above energies E = $10^{19.5}$~eV is less than
 1\%.

\subsection{Performance Boost and Simulation Precision}
\label{performanceBoost}
The individual cosmic ray flux from many sources at large distances to
the observer add up to give an almost isotropic contribution.  The computation
time spent to calculate this background can be eliminated by aborting the
calculation for every individual pixel and adding the total isotropic
background contribution to every pixel. By this we introduce an error
$\epsilon$ to $\vec{p}_{eg}^i$.  To check if the upcoming contributions
are isotropic and decide whether to abort the detailed calculation we
proceed as follows: first we divide the sources into 20 distance bins
and calculate the contribution to $\vec{p}_{eg}^i$ from all sources in
the first bin.  We calculate a factor $a = L_{upc} \cdot
(\max{\vec{p}_{eg}^i} - \min{\vec{p}_{eg}^i}) / (\max{\vec{p}_{eg}^i} +
\min{\vec{p}_{eg}^i}) $ with $L_{upc} = \sum f_S $ being the integrated luminosity
of all sources further away. If $a$ is
lower than a given cut off value, the upcoming luminosity is considered
to be isotropic as the contribution from sources further away is more
isotropic than from nearer sources. The flux from the upcoming bins is
integrated and added once to all pixels in $\vec{p}_{eg}^i$. If $a$
exceeds the selected cut off value we proceed with the sources in the next
bin.